\documentclass[%
 reprint,
superscriptaddress,
groupedaddress,
showpacs,preprintnumbers,
nofootinbib,
nobibnotes,
 amsmath,amssymb,
prb,
]{revtex4-1}

\usepackage{graphicx}
\usepackage{dcolumn}
\usepackage{bm}
\usepackage{amssymb}   
\usepackage{amsmath}
\usepackage{color}
\usepackage[normalem]{ulem}
\usepackage{siunitx}
\usepackage{mathrsfs}
\usepackage{tikz}
\usepackage{amsfonts}
\usepackage{kantlipsum}

\begin{document}


\title{Comparison between the linear and nonlinear homogenization of graphene and silicon metasurfaces}%

\author{Qun Ren}

\author{J. W. You}%

\author{N. C. Panoiu}

\affiliation{%
Department of Electronic and Electrical Engineering, University College London, Torrington Place, London WC1E 7JE, United Kingdom
}%

\date{\today}

\begin{abstract}
In this paper, we use a versatile homogenization approach to model the linear and nonlinear optical
response of two metasurfaces: a plasmonic metasurface consisting of a square array of graphene
cruciform patches and a dielectric metasurface consisting of a rectangular array of photonic
crystal (PhC) cavities in a silicon PhC slab waveguide. The former metasurface is resonant at
wavelengths that are much larger than the graphene elements of the metasurface, whereas the
resonance wavelengths of the latter one are comparable to the size of its resonant components. By
computing and comparing the effective permittivities and nonlinear susceptibilities of the two
metasurfaces, we infer some general principles regarding the conditions under which homogenization
methods of metallic and dielectric metasurfaces are valid. In particular, we show that in the case
of the graphene metasurface the homogenization method describes very well both its linear and
nonlinear optical properties, whereas in the case of the silicon PhC metasurface the homogenization
method is less accurate, especially near the optical resonances.
\end{abstract}

\maketitle

\section{Introduction}\label{Intro}

Metamaterials, whose emergence has opened up exciting new opportunities to create novel media with
pre-designed physical properties, have been proving to have a significant impact on the development
of new approaches and devices for controlling light interaction with matter and achieving key
functionalities, including light focusing \cite{xkpj16nc, gzcql10oe, cz10oe}, perfect lensing
\cite{ss12njp, gm15prl}, perfect absorption \cite{nsjd08prl, sk12prb, rmcf12oe, jsns11oe, sqlf13n},
electromagnetic cloaking \cite{djbs06s, ps09mt, dyygs12nc}, imaging with sub-diffraction resolution
\cite{ag03ieee,dbcn10oe,dtem06prl,zshyycx07oe,cpl08prl}, and optical sensing \cite{obt13prl, apswg09nm, kymem16nm, cjxsn13acs}. One of the most important functionalities provided by
metamaterials is the enhancement of the local optical field \cite{igmha13prb, pegwc11jo, jwzxs14oe,
zarra08m, baj09prb, mdbai14nl, fzp06nl, kwb15acsp}. This feature is particularly relevant to
nonlinear optics since nonlinear optical interactions grow nonlinearly with the applied field.
Promising applications of metamaterials can be found in a broad area of science and engineering,
including optical filters, sensing and infrastructure monitoring, medical devices, remote aerospace
applications, and smart solar power management.

As research in metamaterials advanced, it became clear that the two-dimensional (2D) counterpart of
metamaterials, the so-called metasurfaces, would offer the fastest route to functional devices and
applications. This is so chiefly because most nanofabrication techniques can conveniently be
applied to the planar configuration of metasurfaces. These ultrathin and lightweight optical
devices are generally made of sub-wavelength dielectric or metallic elements arranged in
one-dimensional (1D) or 2D periodic arrays. Equally important from a practical perspective, the
single-layer characteristics of photonic devices based on metasurfaces make them particularly
amenable to system integration. Because of their small thickness, light-matter interaction occurs
in a reduced volume and as such optical losses in metasurfaces are relatively small. Importantly in
nonlinear optics applications, this reduced light propagation distance in metasurfaces means that
phase-matching requirements can be relaxed, which greatly reduces the design constraints of
nonlinear optical devices based on metasurfaces \cite{po07ol, jmjma15prb, ad11ome, cga14prb,
egy16nc,gst17nrm}.

Metasurfaces can mainly be divided into two categories, namely metallic (plasmonic)
\cite{fllfc16sa,ya11prb} and dielectric metasurfaces \cite{yidj14nc,jimed15acs}. Plasmonic
metasurfaces, which exploit the resonant excitation of surface plasmons at specific frequencies
\cite{j15nn,gja11ome,kejtn10prl}, can greatly enhance the local optical field, but this phenomenon
is usually accompanied at optical frequencies by large dissipative losses. Dielectric metasurfaces,
on the other hand, experience much smaller optical losses but only provide limited optical field
enhancement. Moreover, another difference between the two classes of metasurfaces, which is
directly related to the magnitude of the optical losses, is that whereas the resonances in the
plasmonic metasurfaces are relatively broad, the (Mie) resonances associated to dielectric
metasurfaces are particularly narrow. As a result, dielectric metasurfaces are usually much more
dispersive than the plasmonic ones. One effective approach to study metasurfaces is using
homogenization methods, which reduce the metasurface to a homogeneous material with specific linear
and nonlinear retrieved optical coefficients. These effective physical quantities are determined in
such a way that the metasurface and homogenous layer of material have the same optical response.

In this paper, we propose a homogenization method and investigate its accuracy when applied to
plasmonic and dielectric metasurfaces. As plasmonic metasurface we consider a graphene metasurface
\cite{fdf11nl,smt12nm,zkfs15sr} consisting of a square array of free-standing graphene cruciform
patches, whereas the dielectric metasurface is made of a rectangular array of photonic crystal
(PhC) cavities possessing high-$Q$ optical modes, embedded in a silicon PhC slab waveguide. The
rationale for our choice is that the two metasurfaces capture the general characteristics of the
two main classes of metasurfaces, \textit{i.e.} plasmonic and dielectric metasurfaces. In
particular, the cruciform graphene patches possess strong plasmon resonances characterized by
highly confined, enhanced optical near-field. As optical nonlinearity, we consider second-harmonic
generation (SHG) by the nonlocal nonlinear polarization, as symmetry considerations imply that the
SHG by the the local nonlinear polarization is zero \cite{h91BookCh,psl18jo}. Moreover, the silicon
PhC cavities were designed so as to possess two high-$Q$ optical cavity modes separated by the
Raman frequency of silicon, which ensures a strong effective Raman nonlinearity of the metasurface
\cite{ryp18oe}.

The remainder of the paper is organized as follows: In the next section we present the geometrical
configurations and material parameters characterizing the two metasurfaces considered in this work.
Then, in Sec.~\ref{Hom}, we introduce the linear and nonlinear homogenization method used to
extract the constitutive parameters of the metasurfaces, whereas in Sec.~\ref{Res} we apply our
homogenization approach to the two metasurfaces and derive general principles regarding the
conditions in which the predictions of the homogenization method are valid. In particular, we
extract the linear and nonlinear constitutive parameters of the metasurfaces and then compare the
optical response of the metasurfaces with that of their homogenized counterparts. Finally, we
conclude our paper by summarizing the main findings of our study and discussing some of their
implications to future developments pertaining to metamaterials research.

\begin{figure}[!t]
\centering\includegraphics[width=0.45\textwidth]{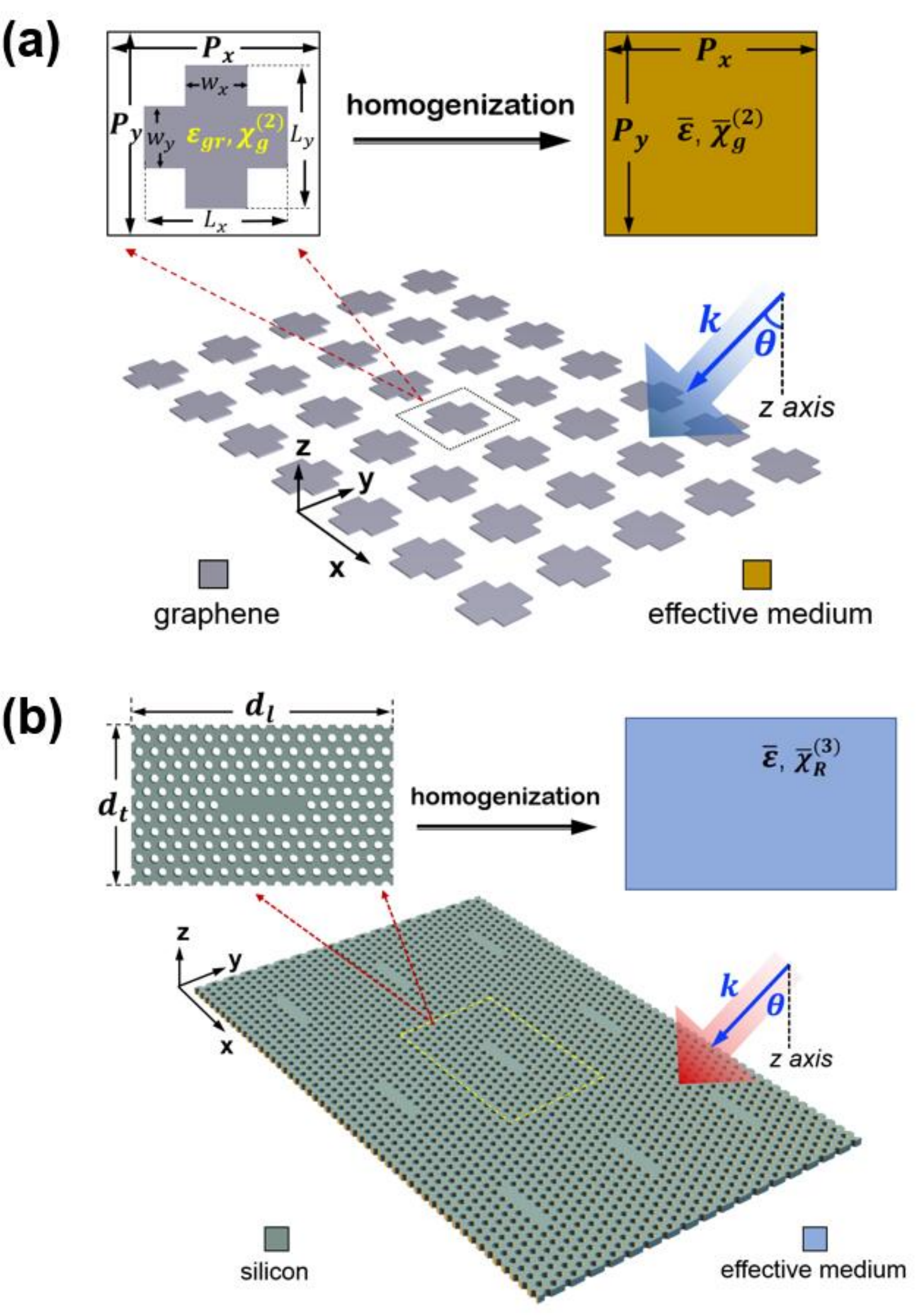} \caption{Schematic of homogenization of two
metasurfaces. (a) Geometry of a graphene metasurface consisting of a 2D array of graphene crosses.
The unit cell is homogenized into a uniform layer of material characterized by effective
parameters. (b) Geometry of a silicon metasurface consisting of a rectangular array of silicon PhC
nanocavities in a hexagonal PhC slab waveguide made of silicon. The unit cell is homogenized into a
uniform layer of material with effective parameters.}\label{config}
\end{figure}

\section{Description of the graphene and silicon metasurfaces}\label{Geom}

In this section, we describe the geometrical configuration and material parameters of the two
metasurfaces investigated in this work, namely the graphene cruciform metasurface illustrated in
figure \ref{config}(a) and the silicon PhC nanocavity metasurface shown in figure \ref{config}(b). In
addition, we explain the rationale for our choice of metasurfaces by presenting their main physical
properties.

\subsection{Geometrical configuration of the graphene and silicon metasurfaces}

The graphene metasurface lies in the $x$-$y$ plane and consists of a square array of cruciform
graphene patches. The symmetry axes of the array coincide with the $x$- and $y$-axes and are
oriented along the arms of the crosses, as per figure \ref{config}(a). The length and width of the
arms of the crosses along the two axes are $L_{x}$ ($L_{y}$) and $w_{x}$ ($w_{y}$), respectively,
whereas the corresponding periods of the metasurface are $P_{x}$ and $P_{y}$. In this, work, unless
otherwise specified, $L_{x}=L_{y}=\SI{60}{\nano\meter}$, $w_{x}=w_{y}=\SI{30}{\nano\meter}$, and
$P_{x}=P_{y}=\SI{100}{\nano\meter}$.

The relative electric permittivity of graphene is given by the relation:
\begin{equation}
\label{eq:eps_g}
\varepsilon_{g}=1+\frac{i\sigma_s}{\varepsilon_0\omega h_{g}},
\end{equation}
where $h_{g}=\SI{0.3}{\nano\meter}$ is the thickness of graphene, $\omega$ is the frequency, and
the graphene surface conductivity, $\sigma_s$, is described by the Kubo's formula \cite{gp16book}:

\begin{eqnarray}\label{eq:sigma}
 \nonumber \sigma_{s} = &\frac{e^2 k_B T\tau}{\pi\hbar^2(1-i\omega\tau)}\left[\frac{\mu_c}{k_B T}+2\ln
\left(e^{-\frac{\mu_c}{k_B T}} + 1\right) \right] \\
&+\frac{ie^2}{4\pi\hbar}\ln\frac{2\vert\mu_c\vert\tau/\hbar-i(1-i\omega\tau)}{2\vert\mu_c\vert\tau/\hbar+i(1-i\omega\tau)}.
\end{eqnarray}
Here, $\mu_c$, $T$, and $\tau$ are the chemical potential, temperature, and relaxation time,
respectively. In this study, we use $\mu_c=\SI{0.6}{\electronvolt}$, $\tau=\SI{0.1}{\pico\second}$,
and $T=\SI{300}{\kelvin}$.

\begin{figure}[!b]
\centering\includegraphics[width=0.45\textwidth]{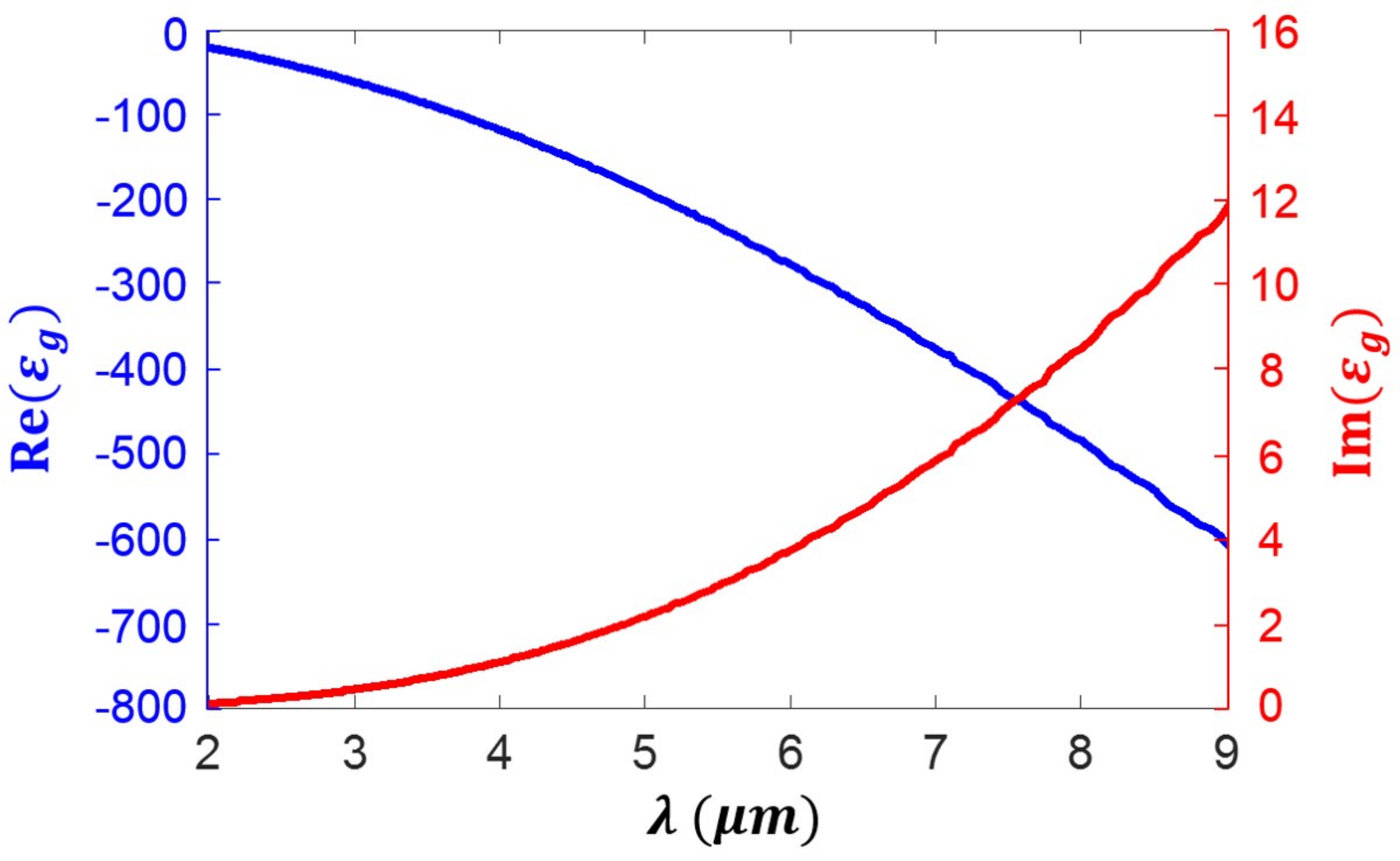} \caption{Wavelength dependence of
the real and imaginary parts of the graphene relative permittivity.} \label{epsil_inrtinsic}
\end{figure}

The wavelength dependence of the real and imaginary parts of graphene permittivity, as described by
Eq.~\eqref{eq:eps_g} in conjunction with Eq.~\eqref{eq:sigma}, are depicted in figure \ref{epsil_inrtinsic}.
It can be seen in this figure that $\mathfrak{Im}(\varepsilon_{g})>0$ and
$\mathfrak{Re}(\varepsilon_{g})<0$, which are spectral characteristics shared by most noble metals.

\begin{figure}[!t]
\centering\includegraphics[width=0.45\textwidth]{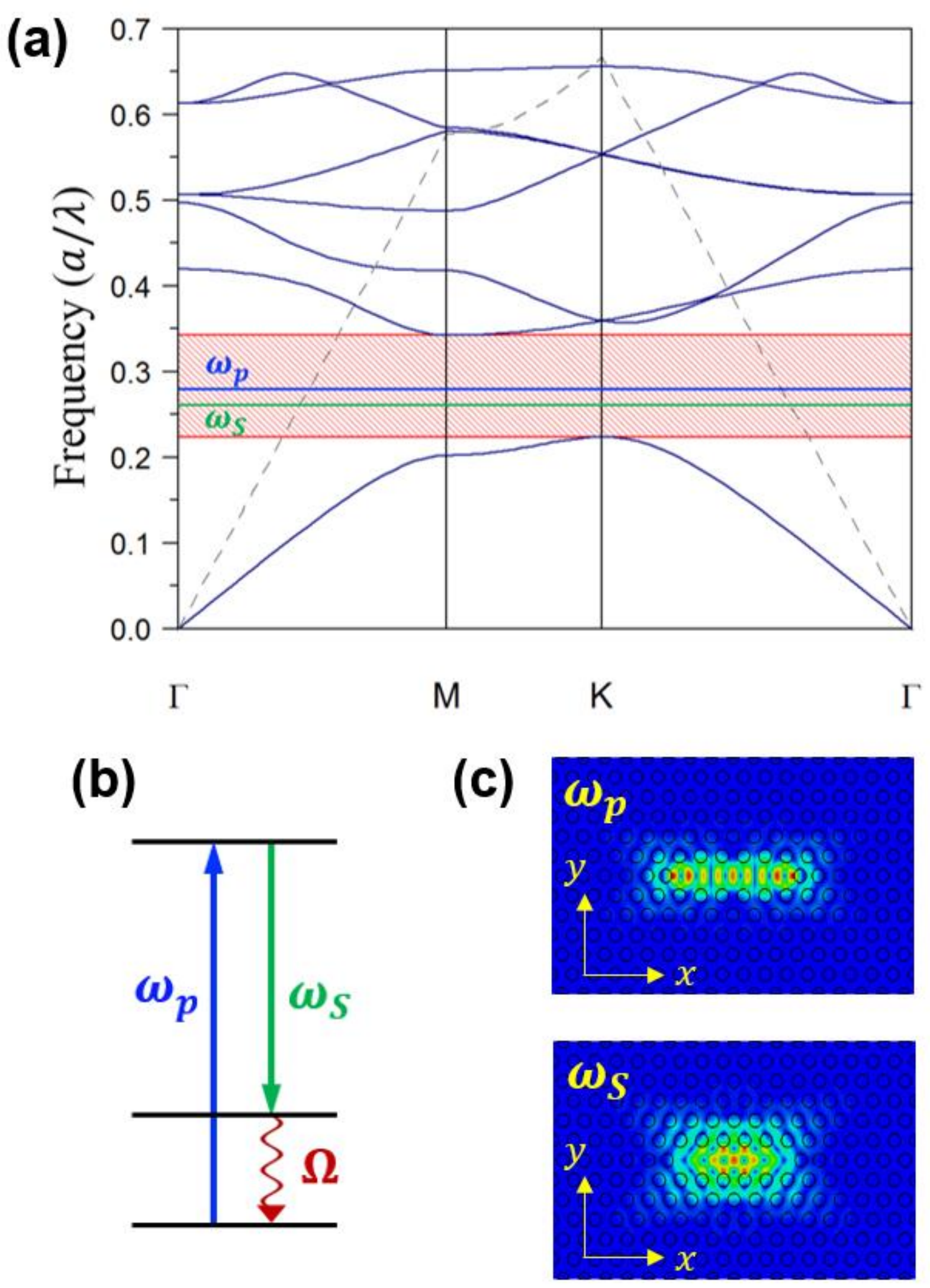} \caption{a) Transverse-magnetic band
structure of the PhC and two optical modes of the PhC cavity withe frequencies of $\omega_{p}$ and
$\omega_{S}$. The dashed lines indicate light lines. b) Diagrammatic representation of the
stimulated Raman scattering. c) The field profiles of the optical modes at the pump and Stokes
frequencies.} \label{band_struct}
\end{figure}

The silicon metasurface, illustrated in figure \ref{config}(b), consists of a rectangular array of
PhC cavities embedded in a PhC slab waveguide made of silicon ($n_{Si}=3.4$) \cite{ryp18oe}. The
PhC slab waveguide comprises a 2D hexagonal lattice of air holes in a silicon slab with the hole
radius $r=0.29a$ and slab thickness $t=0.6a$, where $a$ is the lattice constant. Moreover, the
optical nanocavities are the so-called $L5$ PhC cavities, namely they are created by filling in 5
consecutive holes oriented along the $\Gamma K$ symmetry axis of the hexagonal lattice. The periods
of the PhC metasurface, defined as the center-to-center distance along the $x$- and $y$-axes
between adjacent PhC cavities, are $d_{l}=17a$ and $d_{t}=6\sqrt{3}a$, respectively. In order to
increase the $Q$-factor of the optical modes of this PhC nanocavity, we shifted outwardly the
end-holes of the cavity by $0.15a$ \cite{yw05oe}.

The PhC cavity is designed in such a way that it possesses two optical modes whose frequencies are
separated by the Raman frequency of silicon, $\Omega=2\pi\times\SI{15.6}{\tera\hertz}$
\cite{th73prb}. This ensures a very strong nonlinear coupling between the two optical modes, both
because of the large optical field enhancement inside the cavities and also due to favorable
spatial overlap between the two optical modes. Consequently, one can achieve an efficient Raman
interaction between the two optical modes. This means that the PhC nanocavities can be viewed as
artificially engineered, strongly nonlinear ``silicon meta-atoms'', which when arranged in some
spatial pattern give rise to photonic metasurfaces with large Raman nonlinearity. In particular, if
one chooses the lattice constant $a=\SI{333}{\nano\meter}$, the resonance frequency of the pump and
Stokes modes are $\omega_{p}=\SI{1572.5}{\tera\hertz}$ and $\omega_{S}=\SI{1474.6}{\tera\hertz}$,
respectively, and therefore the condition $\omega_{p}-\omega_{S}=\Omega$ is fulfilled. Expressed in
terms of normalized frequency of $2\pi c/a$, the frequencies of the two optical modes are
$\omega_{p}=0.2778$ and $\omega_{S}=0.2605$. Also, the $Q$-factors of the two modes are
$Q_{p}=1804$ and $Q_{S}=\num{1.12e5}$. Note that the photonic band structure of the PhC slab and
the corresponding cavity modes were computed with RSoft's BandSOLVE \cite{rsoft}, with the cavity
modes lying in the transverse-magnetic bandgap of the PhC slab waveguide (see figure
\ref{band_struct}).
\begin{figure}[!t]
\centering\includegraphics[width=0.45\textwidth]{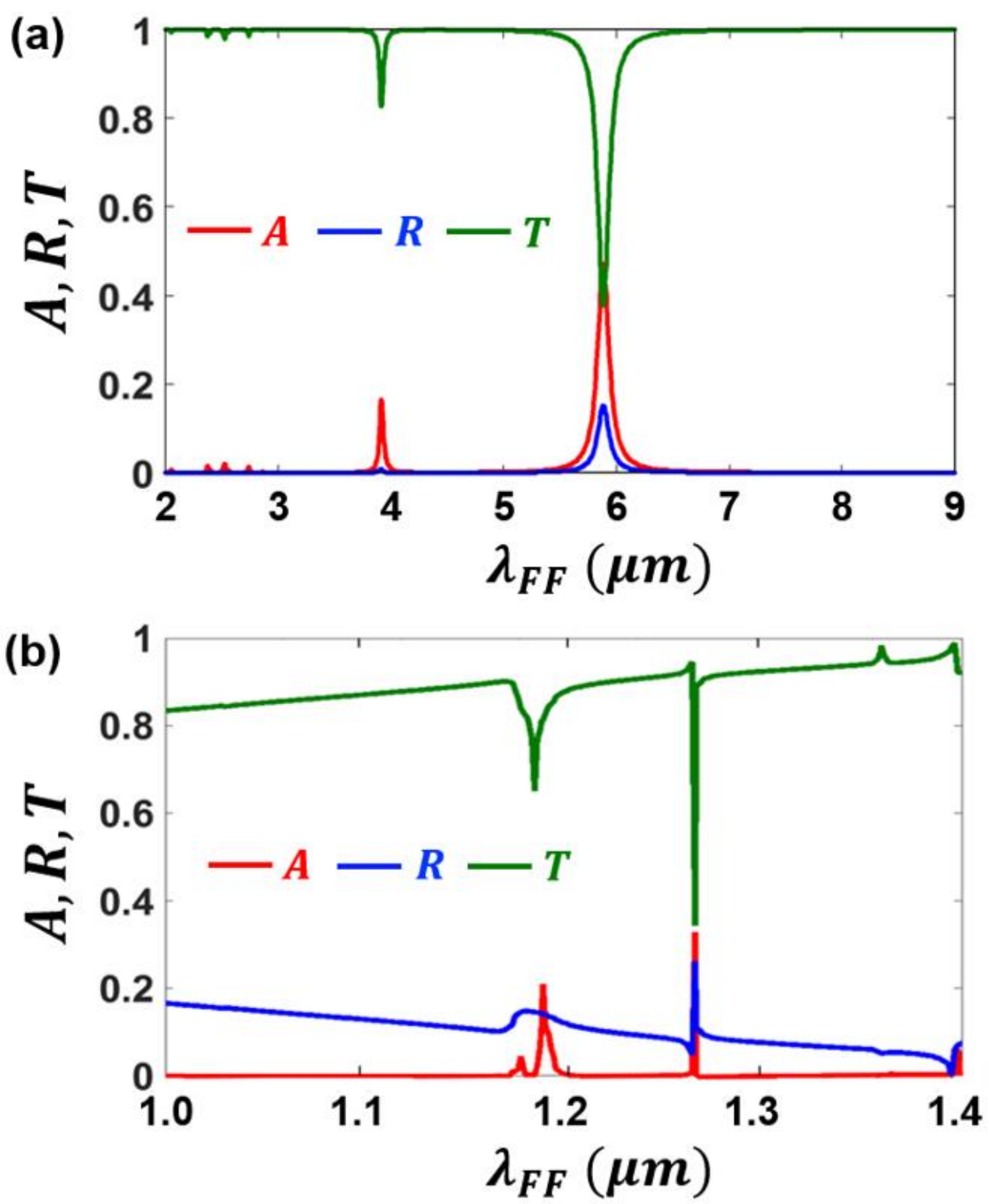} \caption{a) Wavelength dependence of
absorption, reflectance, and transmittance of the graphene metasurface. b) The same as in a), but
determined for the silicon metasurface.} \label{A_R_T}
\end{figure}

The linear optical response of graphene and silicon metasurfaces is presented in figures
\ref{A_R_T}(a) and \ref{A_R_T}(b), respectively, and correspond to an $x$-polarized plane wave
normally incident onto the metasurface. This figure clearly shows that, in the case of the graphene
metasurface, absorption, $A$, reflectance, $R$, and transmittance, $T$, have a series of spectral
resonances occurring at common wavelengths. These resonances are due to the generation of localized
surface plasmons in the graphene crosses, the resonance wavelength of the fundamental plasmon being
$\lambda=\SI{5.88}{\micro\meter}$. Moreover, it can be seen that the excitation of surface plasmons
is accompanied by a large increase of the optical absorption, which suggests that at the
corresponding resonance wavelengths the optical near-field is significantly enhanced. Moreover, the
spectra of the silicon metasurface show two resonances at the frequencies of the two cavity modes,
the width of these resonances being much smaller than that of plasmon resonances.

\section{Linear and nonlinear homogenization method}\label{Hom}

In this section, we present first the homogenization approach used to describe the effective
optical response of the metasurfaces investigated in this paper and retrieve their effective
permittivities and nonlinear susceptibilities and then compare the linear optical response of the
original and homogenized metasurfaces so as to clarify the circumstances in which our
homogenization approach is accurate.

\subsection{Theory of the effective permittivity of metasurfaces}

The general linear homogenization method presented here amounts to establishing a relationship
between the averaged electric displacement field, $\textbf{D}$, and the electric field,
$\textbf{E}$, and is known as the field averaging method. It will be expanded later on to the
nonlinear case. As it is well known, the constitutive relation of a linear anisotropic material is
expressed as $D_{i}=\sum_{j}\epsilon_{ij}E_{j}$, where $\epsilon_{ij}$ is the permittivity tensor
and the subscripts $i,j = x,y,z$. The field-average method relies on this relation, the effective
permittivity of the medium being defined via a similar relation between the electric field and the
electric displacement field:
\begin{eqnarray}
\label{eq:E_average}\overline{\mathbf{E}}(\omega)=&\frac{1}{V}\int_{V}\mathbf{E}(\textbf{r},\omega)
d\textbf{r},\\
\label{eq:D_average}\overline{\mathbf{D}}(\omega)=&\frac{1}{V}\int_{V}\mathbf{D}(\textbf{r},\omega)
d\textbf{r}.
\end{eqnarray}
This method is particularly suitable for describing metasurfaces, because in this case the
averaging domain can be naturally defined as the unit cell of the metasurface. Therefore, in the
equations above, $V$ is the volume of the unit cell of the metasurface.

According to the field averaging method, for an isotropic medium whose permittivity tensor is
diagonal, the effective permittivity is evaluated as
$\overline{\epsilon}_{i}=\overline{D}_{i}/\overline{E}_{i}$. This definition can be extended to
anisotropic media as follows: First, one defines an auxiliary quantity,
$d_{ij}=\epsilon_{ij}E_{j}$, and express the displacement field as $D_{i}=\sum_{j}d_{ij}$, and then
calculate the averaged value of each component of the auxiliary quantity:
\begin{equation}
\label{eq:D_average}
\overline{d}_{ij}(\omega)=\frac{1}{V}\int_{V}d_{ij}(\textbf{r},\omega)d\textbf{r}=\frac{1}{V}\int_{V}\epsilon_{ij}(\textbf{r},\omega)E_{j}(\textbf{r},\omega)d\textbf{r}.
\end{equation}

If we assume that the averaged fields are related by a constitutive relation similar to that
corresponding the local fields, namely
$\overline{D}_{i}=\sum_{j}\overline{\epsilon}_{ij}\overline{E}_{j}$, and by requiring that the
average of the field $\mathbf{D}(\mathbf{r},\omega)$ of the metasurface and the field
$\overline{\mathbf{D}}(\omega)$ in the homogenized layer of material are termwise equal, the
effective permittivity is determined by the following equation:
\begin{equation}
\label{eq:eps_average}
\overline{\epsilon}_{ij}(\omega)=\frac{\overline{d}_{ij}(\omega)}{\overline{E}_{j}(\omega)}=\frac{\displaystyle
\int_{V}\epsilon_{ij}(\textbf{r},\omega)E_{j}(\textbf{r},\omega)d\textbf{r}}{\displaystyle
\int_{V}E_{j}(\textbf{r},\omega)d\textbf{r}}.
\end{equation}
This formula has been used to determine the effective permittivity of both metasurfaces. Before
moving on to the calculation of the effective nonlinear susceptibilities of the metasurfaces, we
note that in the case of the graphene metasurface the volume integrals can be reduced to surface
integrals across the midsection of the graphene sheet because the fields across the ultrathin
graphene layer vary only slightly.

\subsection{Calculation of effective second-order susceptibility of graphene metasurfaces}

As we assume that the graphene crosses are free standing, symmetry considerations imply that the
dipole (local) nonlinear polarization at the second harmonic (SH) exactly cancels. Consequently,
the lowest order SHG is due to the nonlocal nonlinear polarization whose sources are magnetic
dipoles and electric quadrupoles oscillating at the SH frequency. It should be noted that if the
graphene patches lie onto a substrate the inversion symmetry is broken and consequently the
generated SH is due to the dipole nonlinear polarization. The homogenization of such graphene
metasurfaces has been recently studied \cite{ryp19prb}.

This nonlinear polarization and the associated nonlinear surface current in the case of graphene
metasurfaces characterized by nonlocal nonlinear polarization can be expressed as \cite{h91BookCh}:
\begin{eqnarray}
\label{eq:chi_2}
\mathbf{P}(\mathbf{r},\Omega)=\epsilon_{0}\bm{\chi}_{g}^{(2)}(\Omega;\omega,\omega)\vdots
\mathbf{E}(\mathbf{r},\omega)\mathbf{\nabla}\mathbf{E}(\mathbf{r},\omega), \\
\label{eq:sig_2}
\mathbf{J}_{s}(\mathbf{r},\Omega)=\bm{\sigma}_{g}^{(2)}(\Omega;\omega,\omega)\vdots
\mathbf{E}(\mathbf{r},\omega)\mathbf{\nabla}\mathbf{E}(\mathbf{r},\omega),
\end{eqnarray}
where $\bm{\chi}_{g}^{(2)}(\Omega;\omega,\omega)$ and $\bm{\sigma}_{g}^{(2)}(\Omega;\omega,\omega)$
are the \textit{bulk} nonlinear second-order susceptibility and \textit{surface} nonlinear
second-order optical conductivity, respectively, and are related by the following formula:
\begin{equation}
\label{eq:chisig}
\bm{\chi}_{g}^{(2)}(\Omega;\omega)=\frac{i}{\epsilon_{0}\Omega h_{g}}\bm{\sigma}_{g}^{(2)}(\Omega;\omega).
\end{equation}

We stress that instead of a bulk nonlinear susceptibility one can use a surface one, defined as
$\bm{\chi}_{s,g}^{(2)}=h_{g}\bm{\chi}_{g}^{(2)}=(i/\epsilon_{0}\Omega)\bm{\sigma}_{g}^{(2)}$, but
we decided to use bulk quantities so that it is more convenient to compare these nonlinear optical
coefficients of graphene with those of other centrosymmetric materials.

The surface nonlinear second-order optical susceptibility of graphene has been recently derived in
\cite{csa16acsn} and is given by the following equation:
\begin{equation}\label{eq:sig_2_f}
\sigma_{g,ijkl}^{(2)}(\Omega;\omega)=\sigma_{g,\Omega}^{(2)}(\omega)\left(\delta_{ik}\delta_{jl}-\frac{5}{3}\delta_{ij}\delta_{kl}-\frac{1}{3}\delta_{il}\delta_{jk}\right),
\end{equation}
where the scalar part of the surface second-order conductivity tensor is:
\begin{equation}\label{eq:sig_sc}
\sigma_{g,\Omega}^{(2)}(\omega)=\frac{3e^{3}v_{F}^{2}\tau^{3}}{8\pi\hbar^{2}{(1-i\omega\tau)}^{3}}.
\end{equation}

Componentwise, the nonlinear polarization can be evaluated as:
\begin{equation}
\label{eq:Polarization_i}
P_{i}(\mathbf{r},\Omega)=\epsilon_{0}\sum_{jkl}\chi_{g,ijkl}^{(2)}E_{j}(\mathbf{r},\omega)\nabla_{k}E_{l}(\mathbf{r},\omega)\equiv\sum_{jkl}q_{ijkl}
\end{equation}
where we introduced a new nonlinear auxiliary quantity defined as
$q_{ijkl}=\epsilon_{0}\chi_{g,ijkl}^{(2)}E_{j}\nabla_{k}E_{l}$. Moreover, the spatial average of
this quantity is:
\begin{equation}
\label{eq:q_ijk}
\overline{q}_{ijkl}(\Omega)=\frac{1}{V}\int_{V}\epsilon_{0}\chi_{ijkl}^{(2)}(\textbf{r})E_{j}(\textbf{r},\omega)\nabla_{k}
E_{l}(\textbf{r},\omega)d\textbf{r},
\end{equation}
where $\chi_{ijkl}^{(2)}(\textbf{r})=\chi_{g,ijkl}^{(2)}$ if $\textbf{r}$ corresponds to a point
inside the graphene crosses and $\chi_{ijkl}^{(2)}(\textbf{r})=0$ if $\textbf{r}$ is in the air
region.

If we express the nonlinear polarization in the homogenized metasurface as:
\begin{equation}
\label{eq:Polarization_eff}
\overline{P}_{i}(\Omega)=\epsilon_{0}\sum_{jkl}\overline{\chi}_{ijkl}^{(2)}\overline{E}_{j}(\omega)\overline{\nabla_{k}E_{l}}(\omega),
\end{equation}
where $\overline{\chi}_{ijkl}^{(2)}$ is the effective nonlinear second-order susceptibility of the
homogenized metasurface, and impose the condition that the spatial average of the nonlinear
polarization described by Eq.~\eqref{eq:Polarization_i} is \textit{termwise} equal to the polarization in Eq.~\eqref{eq:Polarization_eff}, we obtain the following formula for the effective nonlinear susceptibility:
\begin{equation}
\label{eq:chi_eff}
\overline{\chi}_{ijkl}^{(2)}(\Omega;\omega)=\frac{\overline{q}_{ijkl}}{\overline{E}_{j}\overline{\nabla_{k}E_{l}}}.
\end{equation}
In this formula and in Eq.~\eqref{eq:Polarization_eff}, the quantity $\overline{\nabla_{i}E_{j}}$ is defined as:
\begin{equation}
\label{eq:avnabe}
\overline{\nabla_{i}E_{j}}=\frac{1}{V}\int_{V}\nabla_{i}
E_{j}(\textbf{r},\omega)d\textbf{r}.
\end{equation}

\subsection{Theory of effective Raman susceptibility of silicon metasurfaces}

The calculation of the effective Ramman susceptibility of the silicon based PhC metasurface
described in Sec. 2 is similar to that of the effective second-order susceptibility presented
in the preceding subsection, so that here we present only the main steps. A more detailed
derivation can be found in \cite{ryp18oe}.

We start our analysis with the nonlinear Raman polarization at the Stokes frequency,
$\mathbf{P}_{R}(\mathbf{r},\omega_{S})$, which can be written as:
\begin{equation}
\label{eq:Raman_intr}
\mathbf{P}_{R}(\mathbf{r},\omega_{S})=\frac{3}{2}\epsilon_{0}\bm{\chi}_{R}^{(3)}(\mathbf{r})\vdots
\mathbf{E}(\mathbf{r},\omega_{p})\mathbf{E}^{*}(\mathbf{r},\omega_{p})\mathbf{E}(\mathbf{r},\omega_{S}),
\end{equation}
where $\bm{\chi}_{R}^{(3)}(\mathbf{r})$ is the Raman susceptibility and
$\mathbf{E}(\mathbf{r},\omega_{p})$ and $\mathbf{E}(\mathbf{r},\omega_{S})$ are the optical fields
at the pump and Stokes frequencies, respectively. For the sake of simplicity, we assume that the
symmetry axes of the array of PhC cavities coincide both with the $x$- and $y$-axes and with the
principal axes of silicon. Under these circumstances, the nonzero components of
$\bm{\chi}_{R}^{(3)}$ are
$\chi_{R,ijij}^{(3)}=\chi_{R,jiji}^{(3)}=\chi_{R,jiij}^{(3)}=\chi_{R,ijji}^{(3)}$, with $i,j=x,y,z$
and $i\neq j$. The value at resonance of the only independent component is
$\chi_{R,1212}^{(3)}=-i\SI{11.2e-18}{\square\meter\per\square\volt}$ \cite{jcd04tap}.

We then define the spatially averaged effective Raman polarization:
\begin{equation}
\label{eq:Raman_av}
\overline{\mathbf{P}}_{R}(\omega_{S})=\frac{1}{V}\int_{V}\mathbf{P}_{R}(\mathbf{r},\omega_{S})d\mathbf{r},
\end{equation}
where the volume integration is taken over the unit cell of the metasurface, together with the
effective Raman polarization in a homogenized slab of nonlinear optical medium with the same
thickness as that of the PhC slab:
\begin{equation}
\label{eq:Raman_eff}
\mathbf{P}_{R}^{\mathrm{eff}}(\omega_{S})=\frac{3}{2}\epsilon_{0}\overline{\bm{\chi}}_{R}^{(3)}\vdots \overline{\mathbf{E}}(\omega_{p})\overline{\mathbf{E}}^{*}(\omega_{p})\overline{\mathbf{E}}(\omega_{S}).
\end{equation}
Here, $\overline{\bm{\chi}}_{R}^{(3)}$ is the effective Raman susceptibility of the homogenized
metasurface.

As in the case of the graphene metasurface, we cannot simply impose the condition that the
components of the nonlinear polarizations in Eq.~\eqref{eq:Raman_av} and Eq.~\eqref{eq:Raman_eff} are equal because in the general case the effective Raman susceptibility tensor,
$\overline{\bm{\chi}}_{R}^{(3)}$, has 81 independent components, so that the corresponding system
of equations is overdetermined. Consequently, we impose the condition that the r.h.s. of equations
Eq.~\eqref{eq:Raman_av} and Eq.~\eqref{eq:Raman_eff} are \textit{termwise identical}. Using this constraint,
it can be seen that the components of $\overline{\bm{\chi}}_{R}^{(3)}$ are determined by the
following relations:
\begin{eqnarray}
\label{eq:chi_eff} \overline{\chi}_{R,ijkl}^{(3)}=\frac{\displaystyle
\frac{1}{V}\int_{V}\chi_{R,ijkl}^{(3)}(\mathbf{r}) E_{j}(\mathbf{r},\omega_{p})E_{k}^{*}(\mathbf{r},\omega_{p})E_{l}(\mathbf{r},\omega_{S})d\mathbf{r}}{\displaystyle \overline{E}_{j}(\omega_{p})\overline{E}_{k}^{*}(\omega_{p})\overline{E}_{l}(\omega_{S})}.
\end{eqnarray}
Note that the components of $\overline{\bm{\chi}}_{R}^{(3)}$ and $\bm{\chi}_{R}^{(3)}$ cancel for
the same set of indices $i$, $j$, $k$, and $l$.

\section{Results and Discussions}\label{Res}

In this section, we study the circumstances in which our method produces accurate results and use
it to understand the main differences between the physical properties of graphene and silicon PhC
metasurfaces.

\subsection{Effective permittivities of the graphene and silicon photonic crystal metasurfaces}

\begin{figure}[!b]
\centering\includegraphics[width=0.45\textwidth]{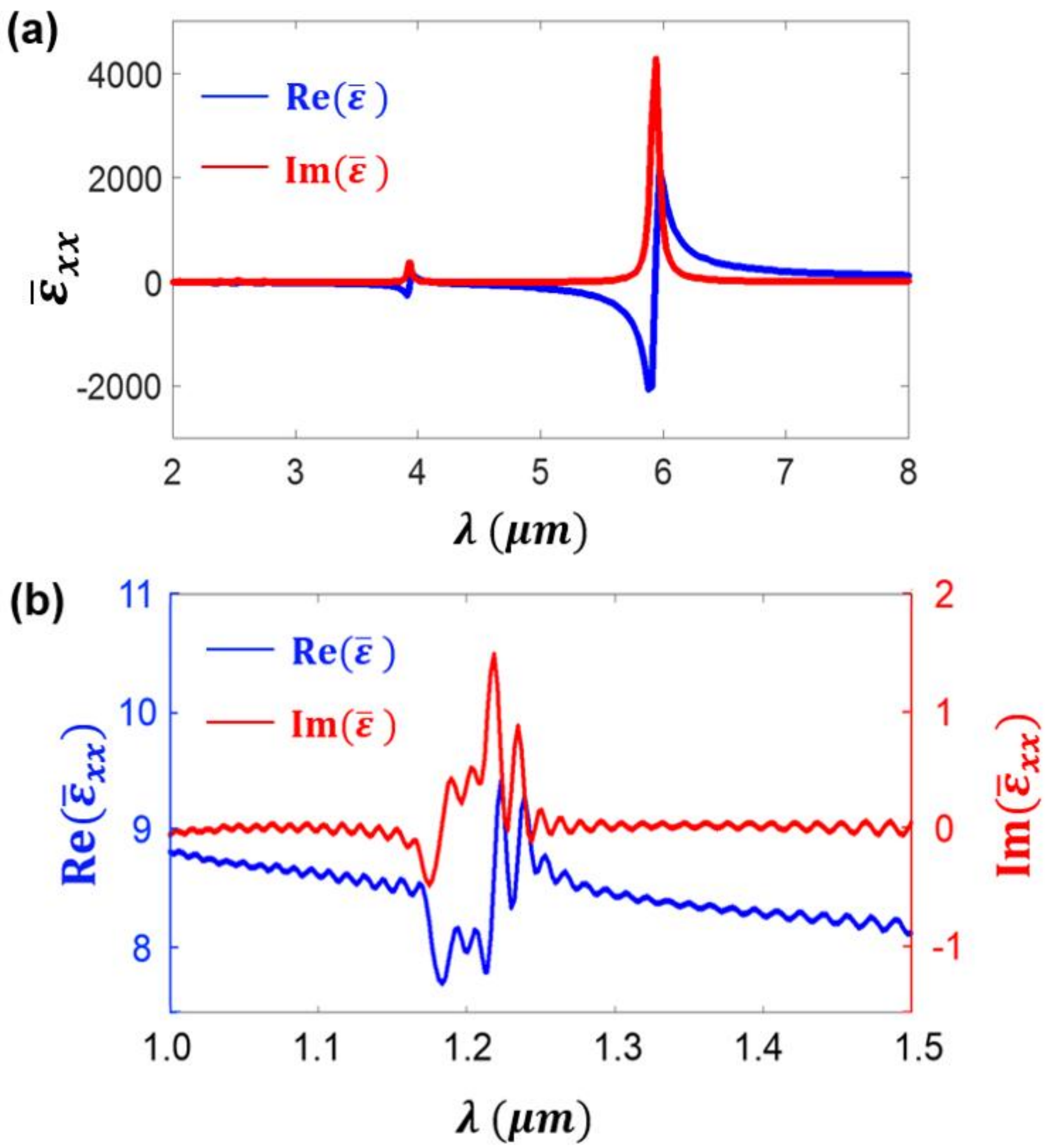}
\caption{a) Spectra of the real and imaginary parts of the effective permittivity of the graphene metasurface. b) The same as in a), but calculated for the silicon metasurface in x-polarized excitation.}
\label{epsil_eff}
\end{figure}
Based on our theoretical analysis, the effective permittivity of the two metasurfaces investigated
in this work can be calculated using Eq.~\eqref{eq:eps_average}. The results of our calculations,
corresponding to the graphene and silicon PhC metasurfaces, are presented in Fig.~\ref{epsil_eff}(a) and Fig.~\ref{epsil_eff}(b), respectively. These figures show some similarities between the two spectra but also significant differences. Thus, the effective permittivity of graphene metasurface displays a series of spectral resonances of Lorentzian nature, which occur at the plasmon resonance wavelengths of the graphene crosses. This means that the graphene crosses behave as meta-atoms that
possess a series of resonant states, the overall optical response of the metasurface being
primarily determined by these resonances. The main reason for this behavior can be traced to the
size of the graphene crosses relative to the resonance wavelengths of the plasmons of the graphene
crosses. Specifically, since the size of the crosses is much smaller than the plasmon wavelengths,
the overall optical response of the graphene metasurface can be viewed as a superposition of the
response of weakly interacting Lorentz-type oscillators.

The spectrum of the effective permittivity of the silicon PhC metasurface, on the other hand,
presents a series of complex features, which are the result of several phenomena. Thus, the two
main resonances of the effective permittivity are due the excitation of the two optical modes of
the PhC cavity. The spectral separation between the frequencies of the two cavity modes is
relatively small and this leads to interference features in the spectrum of the effective
permittivity. The other, weaker spectral peaks are presumably due to leaky modes of the PhC slab.

\subsection{Validation of the homogenization approach}

\begin{figure}[!t]
\centering\includegraphics[width=0.45\textwidth]{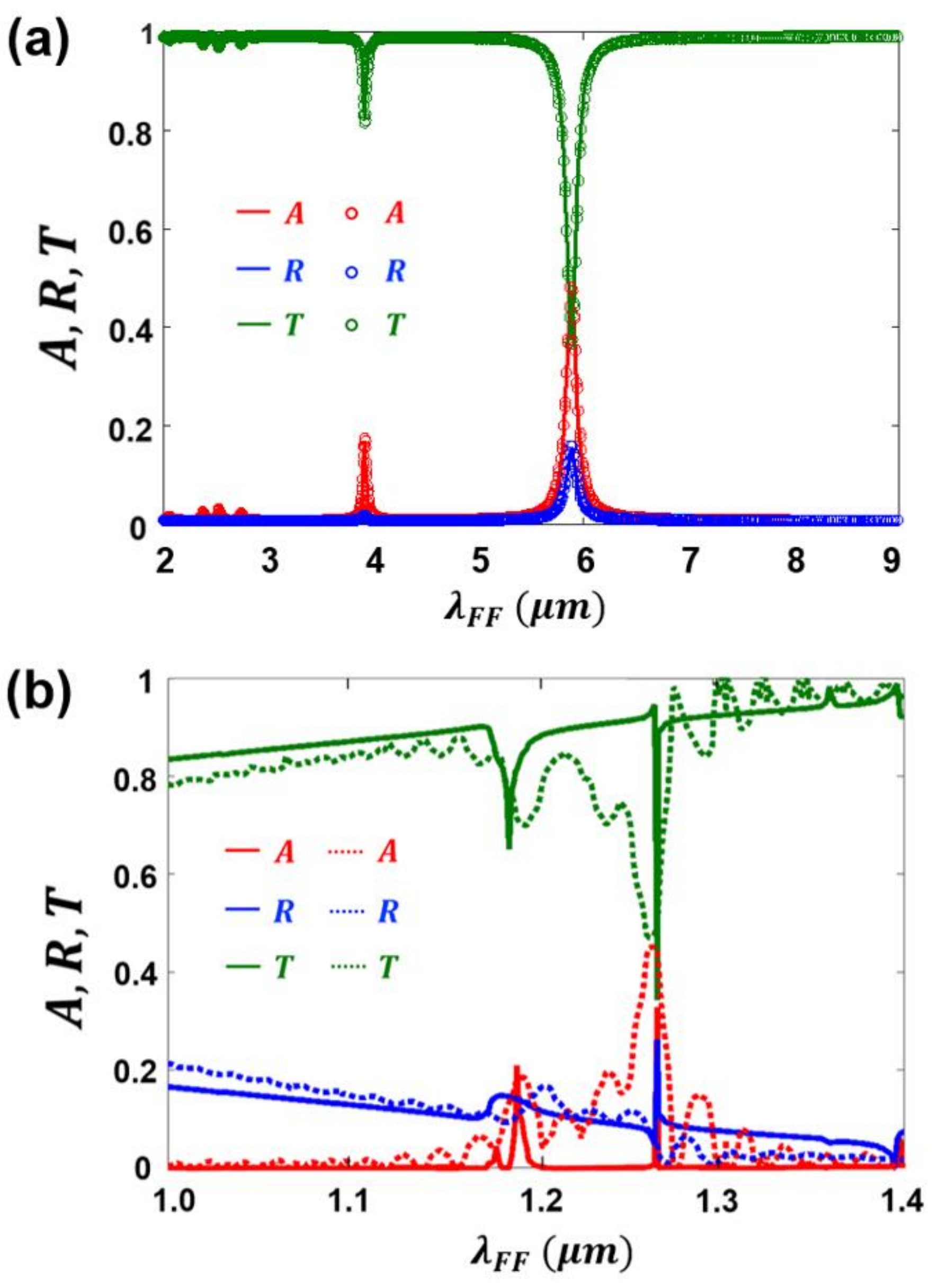}
\caption{a) Wavelength dependence of
absorption, reflectance, and transmittance of the graphene metasurface (solid lines) and its
homogenized counterpart (marked with circles). b) The same as in a), but determined for the silicon
metasurface. The spectra corresponding to the homogenized metasurface are depicted with dotted
lines.}
\label{homo_A_R_T}
\end{figure}
In order to validate the conclusions drawn in the preceding subsection and to investigate the
situations in which our homogenization method is accurate, we calculated the absorption, $A$,
reflectance, $R$, and transmittance, $T$ of both metasurfaces and their homogenized counterparts.
The main results of these calculations are summarized in Fig.~\ref{homo_A_R_T} and they reveal several important ideas. Thus, it can be seen in this figure that in the case of the graphene metasurface the resonances of the transmittance occur at the same wavelengths as the resonances of the effective permittivity, whereas for the silicon PhC metasurface the the two sets of resonances
differ to some extent. In order to explain these results, one should note that generally the
resonances of the transmittance of a planar optical system correspond to excitation of bound modes
of the optical system. In the cases investigated here, the bound modes are localized surface
plasmons of graphene crosses for the graphene metasurface and optical modes of the PhC cavities and
leaky modes of the PhC slab. Moreover, the excitation of these resonances induces a resonant
response of the medium, via the polarization of the medium, which in turn leads to resonances in
the spectrum of the effective permittivity of the homogenized metasurfaces.
\begin{figure}[!b]
\centering\includegraphics[width=0.45\textwidth]{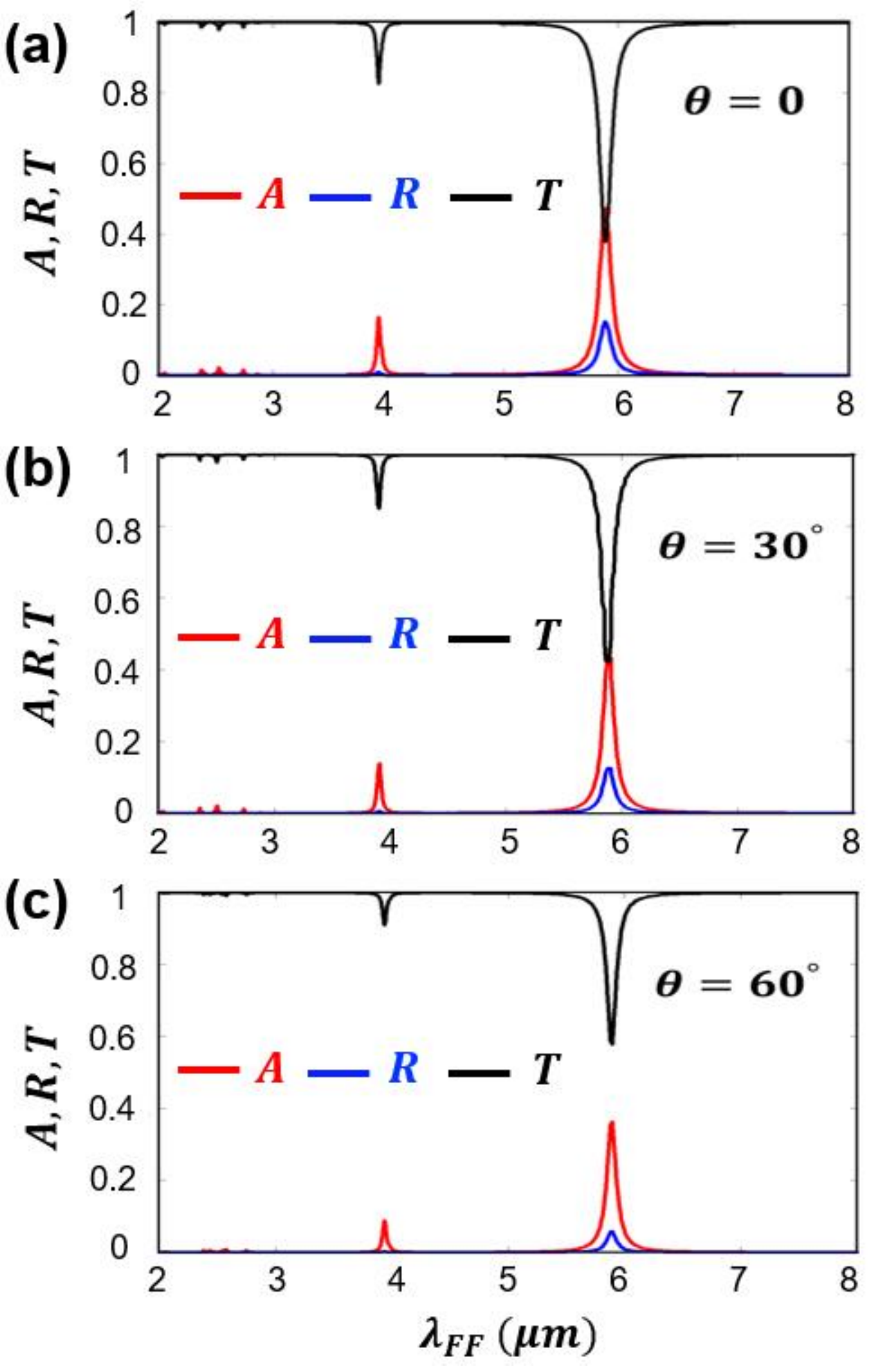} \caption{Spectra of absorption,
reflectance, and transmittance, calculated for different values of the incident angle: a)
$\theta=0$, b) $\theta=30^{\circ}$, and c) $\theta=60^{\circ}$.} \label{A_R_T_theta}
\end{figure}

Fig.~\ref{homo_A_R_T}(b) also shows that whereas the linear optical responses of the graphene
metasurface and its homogenized counterpart are almost identical, in the case of the silicon PhC
metasurface they markedly differ from each other. The main reason for this dichotomy is that the
graphene crosses are much smaller than the operating wavelength, which makes them respond to the
incident optical field as if they were point-like resonators. By contrast, the size of the PhC
cavities is comparable to the resonance wavelength of the cavity modes, which renders our
homogenization method to be less accurate in this case. It should also be noted, however, that
although the homogenization approach for silicon PhC metasurface cannot provide extremely accurate
quantitative values for the effective permittivity, it can still provide us valuable qualitative
insights into the governing physics.

These ideas are further illustrated by the dependence of the linear optical response of the
graphene metasurface on the angle of incidence, $\theta$, of the incoming plane wave, which is
presented in Fig.~\ref{A_R_T_theta}. Thus, it can be observed that the spectral resonances of the
graphene metasurface, calculated for $\theta=$ \SIlist{0;30;60}{\degree}, only slightly varies with
$\theta$, whereas the values of $A$, $R$, and $T$ at the resonance wavelengths depend more
pronouncedly on $\theta$. These findings are explained by the fact that the plasmon resonances
depend chiefly on the shape of the graphene nano-patches, and thus are independent on $\theta$,
whereas the particular values of $A$, $R$, and $T$ depend on the coupling between the incident
field and graphene crosses, more specifically on the spatial overlap between the incident wave and
the optical field of the graphene plasmons, which is obviously $\theta$-dependent.

\subsection{Effective second-harmonic susceptibility of graphene metasurfaces}

\begin{figure}[!b]
\centering\includegraphics[width=0.45\textwidth]{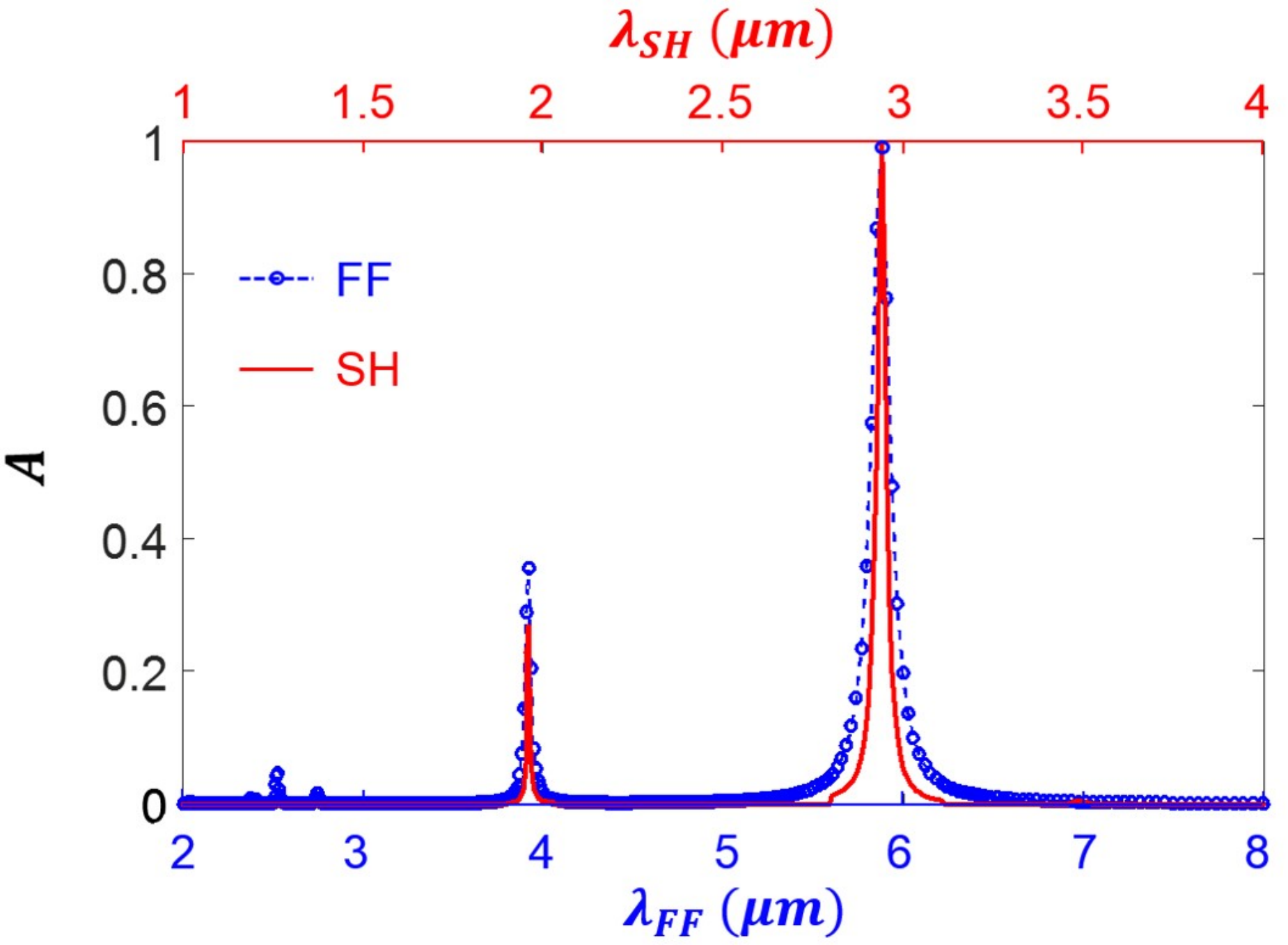}
\caption{Spectrum of optical absorption at FF
(blue curve) and SH (red curve).} \label{SH_FF}
\end{figure}

Let us now consider the nonlinear optical properties of the two optical structures and start with
the graphene metasurface. The excitation of graphene localized plasmons at the fundamental
frequency (FF) induces a strong optical near-field and consequently enhanced nonlinear
polarization, which is the source of the generated SH. This phenomenon can be clearly seen in
Fig.~\ref{SH_FF}, where we present the absorption spectra at the FF and SH. In particular, it can be observed in this figure that the occurrence of a resonance at the FF is accompanied by a resonance at half of its wavelength in the SH spectrum. For example, the plasmon resonance at the FF of $\lambda_{FF}=\SI{5.878}{\micro\meter}$ has a correspondent in the SH spectrum at
$\lambda_{SH}=\lambda_{FF}/2=\SI{2.939}{\micro\meter}$.

\begin{figure}[!t]
\centering\includegraphics[width=0.45\textwidth]{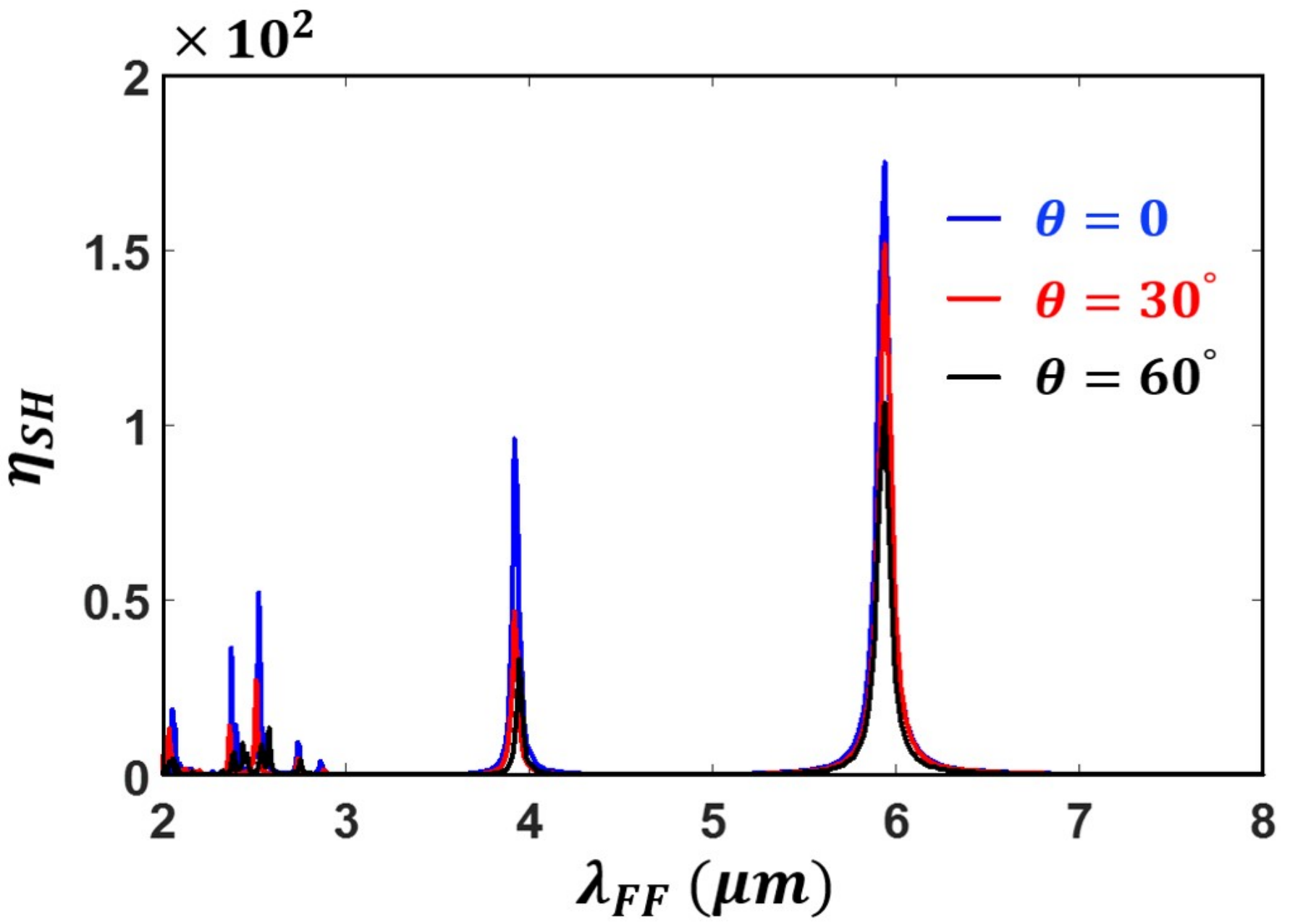}
\caption{Enhancement of effective SH susceptibility of the graphene metasurface, determined for different values of the incident angle, $\theta$. The blue, red, and black curves correspond to $\theta=$ \SIlist{0;30;60}{\degree}, respectively.}
\label{suscept}
\end{figure}

As can be easily inferred from Eq.~\eqref{eq:chisig} and Eq.~\eqref{eq:sig_2_f}, the dominant components (in absolute value) of the second-order susceptibility of graphene are
$\chi_{g,xxyy}^{(2)}=\chi_{g,yyxx}^{(2)}$. The value of this component, determined for a FF
wavelength $\lambda=\SI{1}{\micro\meter}$, is
$\chi_{g,xxyy}^{(2)}=\SI[output-complex-root=\text{\ensuremath{i}}]{-8.37 + 0.133i e-19}{\square\meter\per\volt}$. Moreover, as a consequence of our approach to the calculation of
the effective second-order susceptibility, it is equal to zero for the same set of indices for
which the graphene second-order susceptibility is equal to zero. Therefore, in order to quantify
the enhancement of the nonlinear optical response of the graphene metasurface, we computed the
enhancement factor, $\eta_{SH}=\vert\overline{\chi}_{xxyy}^{(2)}/\chi_{g,xxyy}^{(2)}\vert$, for
several values of the angle of incidence, $\theta=$ \SIlist{0;30;60}{\degree}. We summarize the
results of these calculations in Fig.~\ref{suscept}.

The most important conclusion of this analysis is that, at the wavelength of the fundamental
plasmon, the effective second-order susceptibility of the homogenized graphene metasurface is
enhanced by almost $200\times$. More precisely, the maximum achievable enhancement factor is
$\eta_{SH}=175$, which means that the maximum value of the effective second-order susceptibility of
the homogenized graphene metasurface is
$\vert\overline{\chi}_{xxyy}^{(2)}\vert=\SI{1.46e-16}{\square\meter\per\volt}$. For comparison, the
bulk second-order susceptibility of two centrosymmetric materials widely used in nonlinear optics,
gold and silicon, are $\gamma=\SI{7.13e-21}{\square\meter\per\volt}$ (gold at
$\lambda=\SI{810}{\nano\meter}$) \cite{ktr04jap} and $\gamma=\SI{1.3e-19}{\square\meter\per\volt}$
(silicon at $\lambda=\SI{800}{\nano\meter}$) \cite{fam01ss}. Moreover, it can be inferred from
Fig.~\ref{suscept} that the enhancement factor is smaller for higher-order plasmons, as in this case the plasmon-induced field enhancement decreases. Another feature revealed by Fig.~\ref{suscept} is that the enhancement factor decreases as the angle of incidence increases, a finding explained by the fact that when $\theta$ increases the spatial overlap between the incident wave and the plasmon mode becomes less favorable and thus the enhancement of the local optical field decreases.

\subsection{Effective Raman susceptibility of the silicon photonic crystal metasurface}

Due to the symmetry properties of the silicon PhC metasurface and the orientation of the cavity
array with respect to the principal axes of silicon, the only non-zero component of the effective
Raman susceptibility of the metasurface is
$\overline{\chi}_{R,1212}^{(3)}\equiv\overline{\chi}_{R}^{(3)}$. Therefore, similarly to the case
of the graphene metasurface, we define the enhancement factor
$\eta_{R}=\vert\overline{\chi}_{R}^{(3)}/\chi_{R}^{(3)}\vert$, where
$\chi_{R}^{(3)}\equiv\chi_{R,1212}^{(3)}$ is the dominant component of the Raman susceptibility of
silicon. The parameter $\eta_{R}$ quantifies the enhancement of the Raman nonlinearity of the
silicon PhC metasurface. Moreover, in order to investigate the dependence of the enhancement factor
on the angle of incidence, these calculations were performed for $\theta=$
\SIlist{0;30;60}{\degree}.

Following the procedure we just described, we found out that for the values of the incidence angle
of \SIlist{0;30;60}{\degree}, the enhancement factor was \numlist{2.29e4;3.19e3;1.99e3},
respectively. Thus, it can be seen that a giant enhancement of the effective Raman susceptibility
of the silicon PhC metasurface of more than 4 orders of magnitude can be achieved at normal
incidence. In order to understand the main reason for this remarkable nonlinearity enhancement, one
should note that due to the large $Q$-factor of the pump and Stokes cavity modes the field is
significantly enhanced as compared to the amplitude of the incident wave, which in conjunction with
the fact that the Raman intensity is proportional to the local field to the power of 6, leads to
the extremely large resonant enhancement of the Raman response of the silicon PhC metasurface.
Moreover, the cavity field enhancement decreases when the angle of incidence increases, due to a
weaker coupling between the incoming wave and the cavity modes, which results in reduced
nonlinearity enhancement at larger $\theta$.

As a final remark, it should be noted that the specific design of our metasurface ensures a
particularly efficient Raman amplification. To be more specific, let us compare the spectral width
of Raman interaction in silicon, $\Delta\nu_{R}=\SI{105}{\giga\hertz}$ \cite{opd09aop}, with the
spectral width of the cavity mode at the Stokes frequency, $\Delta\nu_{S}=\omega_{S}/(2\pi
Q_{S})=\SI{2.1}{\giga\hertz}$. Thus, since $\Delta\nu_{S}\ll\Delta\nu_{R}$, an efficient Raman
interaction can be achieved.

\section{Conclusion}\label{Concl}

In summary, two generic metasurfaces, a graphene metasurface based on graphene cruciform patches
and a silicon metasurface with photonic crystal cavities as building blocks, are studied using a
versatile and powerful homogenization method. In particular, in order to quantify the linear and
nonlinear optical response of the two metasurfaces, we computed their effective permittivities and
nonlinear susceptibilities. Our calculations revealed that, in both cases, the nonlinear optical
response of the metasurfaces was enhanced by several orders of magnitude at the resonances of the
metasurface building blocks. Moreover, by comparing the optical response of the metasurfaces and
their homogenized counterparts, we showed that the homogenization approach is more suitable for
graphene-based metasurfaces, because in this case the size of the resonant graphene nanostructures
is much smaller than the operating wavelength. Even though the homogenization approach for
silicon-based metasurfaces appeared to be less accurate, it could still provide valuable
qualitative insights into their nonlinear optical response.

It should be noted that our homogenization approach is rather general, in that it can be readily
extended to metasurfaces of different configurations or made of optical media with various
dispersive and nonlinear optical properties. Moreover, the ideas presented in this paper have broad
applicability, as they can be easily extended to other nonlinear optical interactions of practical
interest, including third-harmonic generation, four-wave mixing, and sum- and difference frequency
generation.

\section*{Funding}

European Research Council (ERC) (ERC-2014-CoG-648328); China Scholarship Council (CSC) and
University College London (UCL) (201506250086).

\end{document}